\RequirePackage{ifpdf}
\ifpdf 
\documentclass[pdftex]{sigma}
\else
\documentclass{sigma}
\fi

\usepackage[thinlines,thicklines]{easybmat}

\begin{document}

\allowdisplaybreaks

\renewcommand{\PaperNumber}{072}

\FirstPageHeading

\ShortArticleName{Clif\/ford Fibrations and Possible Kinematics}

\ArticleName{Clif\/ford Fibrations and Possible Kinematics}

\Author{Alan S. MCRAE}

\AuthorNameForHeading{A.S.~McRae}

\Address{Department of Mathematics, Washington and Lee University, Lexington, VA  24450-0303, USA }
\Email{\href{mailto:mcraea@wlu.edu}{mcraea@wlu.edu}}

\ArticleDates{Received April 10, 2009, in f\/inal form June 19, 2009;  Published online July 14, 2009}

\Abstract{Following Herranz and Santander~[Herranz F.J., Santander M., {\it Mem.\ Real Acad.\ Cienc. Exact.\ Fis.\ Natur. Madrid} {\bf 32} (1998), 59--84, physics/9702030] we will construct homogeneous spaces based on possible kinematical algebras and groups~[Bacry~H., Levy-Leblond~J.-M., {\it J. Math. Phys.} {\bf 9} (1967), 1605--1614] and their contractions for 2-dimensional spacetimes.  Our construction is dif\/ferent in that it is based on a generalized Clif\/ford f\/ibration:  Following Penrose [Penrose R.,
 Alfred A.~Knopf, Inc., New York, 2005] we will call our f\/ibration a Clif\/ford f\/ibration and not a~Hopf f\/ibration, as our f\/ibration is a~geometrical construction.  The simple algebraic properties of the f\/ibration describe the geometrical properties of the kinematical algebras and groups as well as the spacetimes that are derived from them.  We develop an algebraic framework that handles all possible kinematic algebras save one, the static algebra.}


\Classification{11E88; 15A66; 53A17}

\newcommand{\ka}{{\kappa_1}}
\newcommand{\kb}{{\kappa_2}}
\newcommand{\kc}{{\kappa_1 \kappa_2}}
\newcommand{\ga}{{\gamma_1}}
\newcommand{\gb}{{\gamma_2}}
\newcommand{\gc}{{\gamma_3}}
\newcommand{\Ka}{{K_1}}
\newcommand{\Kb}{{K_2}}
\newcommand{\Q}{{\mathbb{H}_{\ka, \kb}}}
\newcommand{\R}{{\mathbb{R}}}
\newcommand{\C}{{\mathbb{C}}}
\newcommand{\I}{{\mathbf{i}}}
\newcommand{\J}{{\mathbf{j}}}
\newcommand{\K}{{\mathbf{k}}}
\newcommand{\Ca}{{\C_{\ka}}}
\newcommand{\Cb}{{\C_{\kb}}}
\newcommand{\Cab}{{\C_{\ka \kb}}}
\newcommand{\SO}{{SO_{\ka,\kb}(3)}}
\newcommand{\so}{{so_{\ka,\kb}(3)}}
\newcommand{\Ck}{{C_{\kappa}}}
\newcommand{\Sk}{{S_{\kappa}}}
\newcommand{\CA}{{C_{\ka}}}
\newcommand{\CB}{{C_{\kb}}}
\newcommand{\CC}{{C_{\ka \kb}}}
\newcommand{\SA}{{S_{\ka}}}
\newcommand{\SB}{{S_{\kb}}}
\newcommand{\SC}{{S_{\ka \kb}}}
\newcommand{\Tk}{{T_{\kappa}}}
\newcommand{\ns}{{\hat{n} \cdot \vec{\gamma}}}

\begin{flushright}
\begin{minipage}{14cm}\it
\noindent As long as algebra and geometry have been separated, their progress have been slow and their uses limited; but when these two sciences have been united, they have lent each mutual forces, and have marched together towards perfection.\\[1mm]
\null \hfill  Joseph Louis Lagrange (1736--1813)
\end{minipage}
\end{flushright}


The nice role played by quaternions in describing rotations of Euclidean 3-dimensional space, and the beauty of the Hopf f\/ibration of the 3-sphere, can be simply generalized for the study of $(1 + 1)$ kinematics.  It is the purpose of this paper to show how this can be done.   We can let $\I$, $\J$, and $\K$ denote the basis of the imaginary part of a generalized quaternion number system so that they also describe a basis for any given kinematical algebra save the static algebra.  The space of unit quaternions (under a suitable choice of norm) then describes a ``3-sphere''.   If $q$ is a point on this sphere, then the Hopf f\/lows $\I q$, $\J q$, and $\K q$ describe f\/ibrations of the sphere where the base spaces are the space of events, the space of space-like geodesics, or the space of time-like geodesics.  The description given below is of a unif\/ied approach to all kinematical algebras (save the state algebra) as well as to the three classical Riemannian surfaces of constant curvature.

\section{Possible kinematics}

It is the purpose of this section to brief\/ly review Bacry and L\'{e}vy-Leblond's work on possible kinematics.  Bacry and L\'{e}vy-Leblond's investigations into the nature of all possible Lie algebras for kinematical groups given three basic principles
\begin{itemize}\itemsep=0pt
\item[(i)] space is isotropic and spacetime is homogeneous;
\item[(ii)] parity and time-reversal are automorphisms of the kinematical group;
\item[(iii)] the one-dimensional subgroups generated by the boosts are non-compact
\end{itemize}
gave rise to 11 possible kinematical algebras.  Restricting our attention to 2-dimensional spacetimes we still obtain the same 11 kinds of algebras (see~\cite{M07}), where each of the kinematical groups is generated by its inertial transformations as well as its spacetime translations.  These groups consist of the de Sitter groups and their contractions.  The physical nature of a contracted group is determined by the nature of the contraction itself, along with the nature of the parent de Sitter group.  The names of the 2-dimensional groups are given in Table~\ref{table1}.
\begin{table}[t]
\centering
  \caption{The 11 possible kinematical groups.}\label{table1}
  \vspace{1mm}

 \begin{tabular}{ c | c  }
\hline
Symbol &  Name  \\
 \hline
\tsep{1mm} $dS$               & de Sitter groups \\
$adS$                 & anti-de Sitter groups \\
$M$                   & Minkowski groups \\
$M_+$     & expanding Minkowski groups\\
$M^{\prime}$              & para-Minkowski groups\\
$C$                    & Carroll groups\\
$N_+$               & expanding Newtonian Universe groups \\
$N_-$                & oscillating Newtonian Universe groups \\
$G$                   & Galilei group \\
$SdS$               & static de Sitter Universe groups \\
$St$                   & static Universe group
 \end{tabular}
 \end{table}

In this paper we will restrict our attention to 2-dimensional spacetimes.  So let $K$ denote the generator of the inertial transformations, $H$ the generator of time translations, and $P$ the gene\-ra\-tor of space translations.  The kinematical algebras are determined by the structure constants~$p$,~$h$, and~$k$ that are given by the commutators
\[
 \left[ K, H \right] = pP,  \qquad \left[ K, P \right] = h H, \qquad \mbox{and}  \qquad \left[ H, P \right] = k K.
\]
If we normalize the structure constants to lie in the set $\{ -1, 0, 1 \}$, then the characteristic Lie brackets for the kinematical Lie algebras are as given in Table~\ref{table2} (see~\cite{BL67}).
\begin{table}[t]\centering
 \caption{The characteristic Lie brackets for the kinematical Lie algebras.}\label{table2}
 \vspace{1mm}

  \begin{tabular}{ c | c | c | c | c | c | c | c | c | c | c | c } \hline \hline
 &$dS$&$adS$&$M$&$M_+$&$M^{\prime}$&$C$&$N_+$&$N_-$&$G$&$SdS$&$St$  \\ \hline
 $[H,P]$&$-K$ &$K$&0&$-K$ &$K$&0 &$-K$ &$K$ &0 &$-K$ &0 \\
 $[K,H]$&$P$& $P$& $P$&0 &0 &0 &$P$ &$P$ &$P$ &0 &0 \\
 $[K,P]$&$H$& $H$& $H$&$H$ &$H$ &$H$ &0 &0 &0 &0 &0
 \\ \hline \hline
 \end{tabular}
 \end{table}

\begin{figure}[t]
\centering
\includegraphics{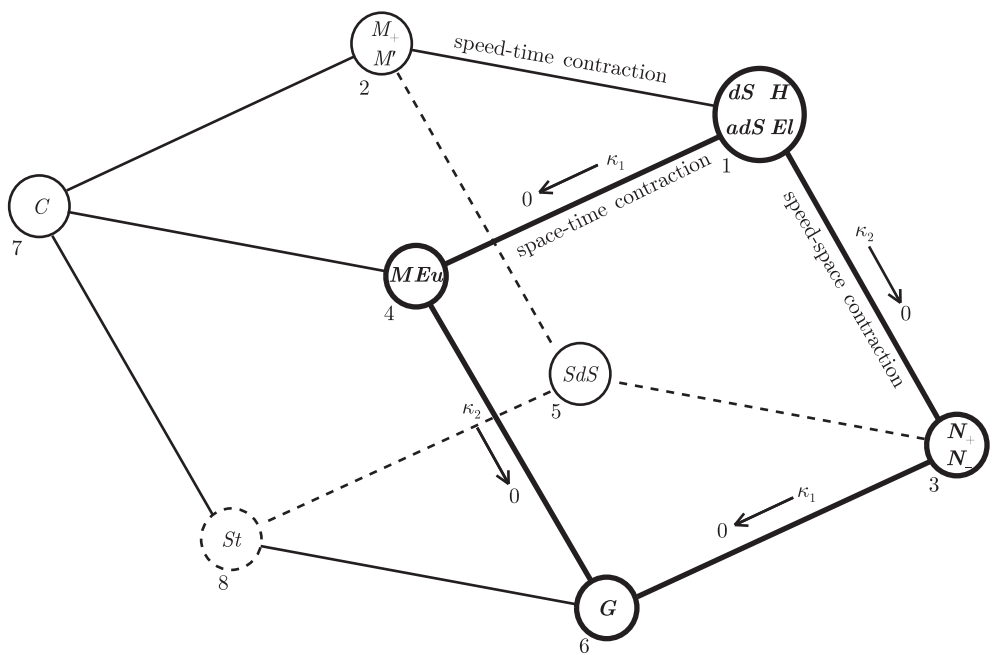}
\caption{The 11 kinematical and 3 non-kinematical groups.}\label{Fig1}
\end{figure}

We will follow Herranz, Ortega and Santander (see~\cite{HS98}) and reduce the number of structure constants from three to two as follows.  The kinematical algebras $dS$, $adS$, $M$, $N_+$, $N_-$, and~$G$ (after rescaling) are determined by the structure constants $\ka$ and $\kb$ that are given by the commutators
\begin{gather}\label{eq1}
 \left[ K, H \right] = P,  \qquad \left[ K, P \right] = -\kb H, \qquad \mbox{and}  \qquad \left[ H, P \right] = \ka K.
 \end{gather}
The constant $\ka = \pm \frac{1}{\tau^2}$ gives the spacetime curvature $\ka$ as well as the universe (time) radius~$\tau$, and the constant $\kb = -\frac{1}{c^2}$ gives the speed of light\footnote{See~\cite{HOS00}.  We will also demonstrate that $\kb = -\frac{1}{c^2}$ and that $\ka$ is the spacetime curvature later on in this paper.} $c$.  For the de Sitter groups $\ka < 0$ and $\kb < 0$, while for the anti-de Sitter groups $\ka > 0$ and $\kb < 0$.  The remaining kinematical algebras (save for $St$) can be obtained by group contractions ($\ka \rightarrow 0$ or $\kb \rightarrow 0$) in possible conjunction with the symmetries $S_P$, $S_H$, and $S_K$:
\begin{gather*}
 S_P : \ \{ K \leftrightarrow H \}: \qquad \left[ K, H \right] = - P,  \qquad \left[ K, P \right] = \ka H, \qquad \mbox{and}  \qquad \left[ H, P \right] = -\kb K,\\
 S_H : \ \{ K \leftrightarrow P \}: \qquad \left[ K, H \right] = -\ka P,  \qquad \left[ K, P \right] = \kb H, \qquad \mbox{and}  \qquad \left[ H, P \right] = - K
\end{gather*}
and
\begin{gather*}
 S_K : \ \{ H \leftrightarrow P \}: \qquad \left[ K, H \right] =  -\kb P,  \qquad \left[ K, P \right] =  H, \qquad \mbox{and}  \qquad \left[ H, P \right] = - \ka K .
\end{gather*}
See Fig.~\ref{Fig1} for an illustration of how the dif\/ferent groups are related via contractions and symmetries:  $El$, $Eu$, and $H$ denote the (non-kinematical) isometry groups of the elliptical, Euclidean, and hyperbolic planes (of constant curvature $\ka$) respectively (see Table~\ref{table3}).

\begin{table}[t]
\centering
 \caption{The characteristic Lie brackets for the non-kinematical Lie algebras.}\label{table3}
 \vspace{1mm}

 \begin{tabular}{ c | c | c | c  } \hline \hline
 & $El$ & $Eu$ & $H$  \\ \hline
 $[H,P]$& $K$ & 0 & $-K$ \\
 $[K,H]$& $P$ & $P$ & $P$ \\
 $[K,P]$& $-H$ & $-H$ & $-H$
 \\ \hline \hline
 \end{tabular}
 \end{table}

For example, if $\kb \rightarrow 0$ then $dS$ and $H$ contract to $N_+$
while $adS$ and $El$ contract to $N_-$, while if $\ka \rightarrow 0$ then $dS$ and $adS$ contract to $M$ while $H$ and $El$ contract to $Eu$.  Similarly a~space-time contraction sends either $M_+$ or $M^{\prime}$ to $C$.  In this paper we will specif\/ically work with the nine kinds of groups indicated in bold in Fig.~\ref{Fig1}, since the other groups (save $St$) are then easily obtained from these nine: the lie algebras of our nine groups have the commutators as given by~\eqref{eq1}.  Henceforward we will not refer to the other algebras.

We can contract with respect to any subgroup, giving us three fundamental types of contraction: {\it speed-space, speed-time}, and {\it space-time contractions}, corresponding respectively to contracting to the subgroups generated by $H$, $P$, and $K$.

\begin{table}[t]\centering
 \caption{The 3 basic symmetries are given as ref\/lections of Fig.~\ref{Fig1}.}
 \vspace{1mm}

 \begin{tabular}{ c | c | c} \hline
 Symmetry & Ref\/lection across face & Corresponding group transformations \\ \hline \hline
 $S_H$ & $1378$ & $M \longleftrightarrow M^{\prime}$, $Eu \longleftrightarrow M_+$, $G \longleftrightarrow SdS$ \\
 $S_P$ & $1268$ & $C \longleftrightarrow SdS$, $M \longleftrightarrow N_+$, $Eu \longleftrightarrow N_-$\\
 $S_K$ & $1458$ & $C \longleftrightarrow G$, $M_+ \longleftrightarrow N_-$, $M^{\prime} \longleftrightarrow N_+$\\ \hline
 \end{tabular}

 \end{table}

 \begin{table}[t]\centering
\caption{Important classes of kinematical groups and their geometrical conf\/igurations in Fig.~\ref{Fig1}.}
\vspace{1mm}

 \begin{tabular}{ l | l } \hline
 Class of groups & Face \\ \hline \hline
 Relative-time & $1247$ \\
 Absolute-time & $3568$ \\
 Relative-space & $1346$ \\
 Absolute-space & $2578$ \\
 Cosmological & $1235$ \\
 Local & $4678$ \\ \hline
 \end{tabular}
  \end{table}

{\it Speed-space contractions.}   We make the substitutions $K \rightarrow \epsilon K$ and $P \rightarrow \epsilon P$ into the Lie algebra and then calculate the singular limit of the Lie brackets as $\epsilon \rightarrow 0$.  Physically the velocities are small when compared to the speed of light, and the spacelike intervals are small when compared to the timelike intervals.  Geometrically we are describing spacetime near a~timelike geodesic, as we are contracting to the subgroup that leaves this worldline invariant, and so are passing from relativistic to absolute time.

{\it Speed-time contractions.}  We make the substitutions $K \rightarrow \epsilon K$ and $H \rightarrow \epsilon H$ into the Lie algebra and then calculate the singular limit of the Lie brackets as $\epsilon \rightarrow 0$.  Physically the velocities are small when compared to the speed of light, and the timelike intervals are small when compared to the spacelike intervals.  Geometrically we are describing spacetime near a spacelike geodesic, as we are contracting to the subgroup that leaves invariant this set of simultaneous events, and so are passing from relativistic to absolute space.  Such a spacetime may be of limited physical interest, as we are only considering intervals connecting events that are not causally related.

{\it Space-time contractions.}  We make the substitutions $P \rightarrow \epsilon P$ and $H \rightarrow \epsilon H$ into the Lie algebra and then calculate the singular limit of the Lie brackets as $\epsilon \rightarrow 0$.  Physically the spacelike and timelike intervals are small, but the boosts are not restricted.  Geometrically we are describing spacetime near an event, as we are contracting to the subgroup that leaves invariant only this one event, and so we call the corresponding kinematical group a {\it local group} as opposed to a {\it cosmological group}.

\section{Generalized complex numbers}

The generalized complex numbers are not new to physics or mathematics (see~\cite{Y79} for example).  It is the purpose of this section to introduce these numbers to the reader who is not already familiar with them.

\begin{definition}
By the complex number plane $\C_{\kappa}$ we will mean the set of numbers of the form $\{ z = x + i y \; | \; (x,y) \in \R^2 \  \mbox{and} \  i^2 = -\kappa \}$, where the constant $\kappa$ is real and $i$ is not.  $\C_{\kappa}$ is a~real commutative algebra and also has zero divisors when $\kappa \leq 0$.  The real part of $z$ is given by $\mathcal{R}(z) = x$ and the imaginary part by $\mathcal{I}(z) = y$.
\end{definition}

Zero divisors play an important role in determining the conformal structure of spacetime, and although it does not make good algebraic sense to divide by them, one can form the Riemann sphere $\Sigma_{\kappa}$.

\begin{definition}  Let $\Sigma_{\kappa}$ denote the Riemann sphere consisting of the set of all equivalence classes $\left[ \frac{A}{B} \right]$ of complex ratios $\frac{A}{B}$, where $A, B \in \C_{\kappa}$ and where either $A$ or $B$ is not a zero-divisor:  $\frac{A}{B} \sim \frac{C}{D} \Longleftrightarrow A = \mu C$ and $B = \mu D$ for some $\mu \in \C_{\kappa}$ where $\mu$ is not a zero divisor.
\end{definition}

\begin{figure}[t]
\centering
\includegraphics[scale=0.80]{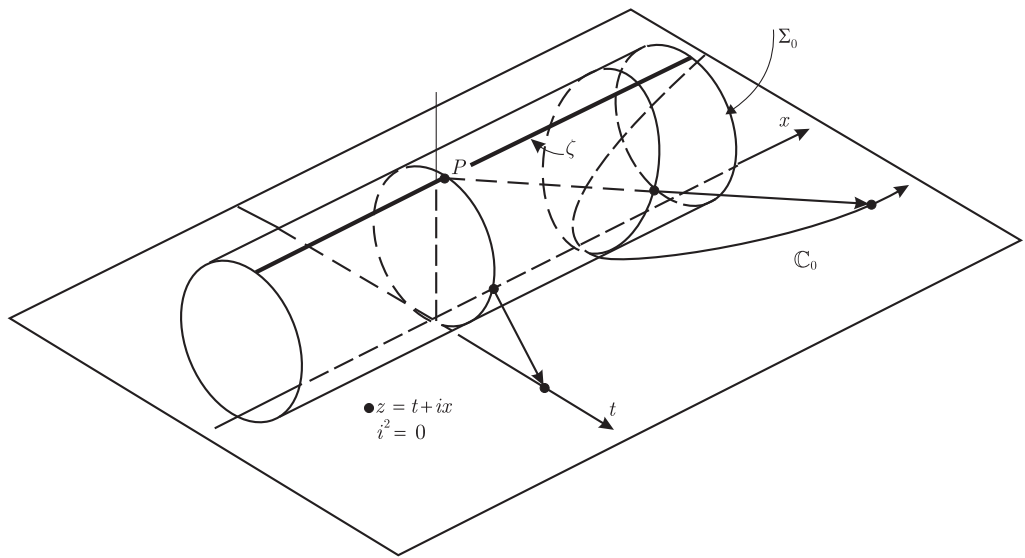}
\caption{The Riemann sphere $\Sigma_0$.}\label{Fig2}
\end{figure}

We can describe $\Sigma_{\kappa}$ through stereographic projection, giving a circular cylinder when $\kappa = 0$ and a hyperboloid of one sheet when $\kappa < 0$ (see \cite{Y79} for details).  Fig.~\ref{Fig2} shows such a~construction for $\Sigma_0$:  Here we are projecting from the point~$P$ onto the complex number plane~$\C_0$ (where the zero divisors consists of all purely imaginary numbers), so that numbers of the form~$\frac{1}{ai}$ correspond to the line~$\zeta$ of ``inf\/inities'' on $\Sigma_0$.  So $P = \left[ \frac{1}{0} \right]$, for example.

The unit circle $z \overline{z} = x^2 + \kappa y^2 = 1$ in $\C_{\kappa}$, where $\overline{z} = x - iy$, is determined by the Hermitian metric $dz d\overline{z} = dx^2 + \kappa dy^2$.  The unit circle can be used to def\/ine the cosine
\begin{equation*}
\Ck(\phi) =
\begin{cases}
\cos{\left(\sqrt{\kappa} \, \phi \right) },                   &\text{if $\kappa  > 0$},   \\
1,  &\text{if $\kappa = 0$},      \\
\cosh{\left( \sqrt{-\kappa} \, \phi \right) },   &\text{if $\kappa < 0$}           \\
\end{cases}
\end{equation*}
and sine
\begin{equation*}
\Sk(\phi) =
\begin{cases}
\frac{1}{\sqrt{\kappa}}\sin{\left( \sqrt{\kappa} \, \phi \right) },                   &\text{if $\kappa > 0$},   \\
\phi,  &\text{if $\kappa = 0$},      \\
\frac{1}{\sqrt{-\kappa}} \sinh{\left( \sqrt{-\kappa} \, \phi \right) },   &\text{if $\kappa < 0$}           \\
\end{cases}
\end{equation*}
functions.  Here $e^{i \phi} = \Ck(\phi) + i\Sk(\phi)$ is a point on the connected component of the unit circle containing $1$, and $\phi$ is the signed distance from $1$ to $e^{i\phi}$ along the circular arc, def\/ined modulo the length $\frac{2\pi}{\sqrt{\kappa}}$ of the unit circle when $\kappa > 0$\footnote{When $\kappa = 0$ the distance along the unit circle $x^2 = 1$ is def\/ined by $ds^2 = dy^2$, as the Hermitian metric $ds^2 = dx^2 + \kappa dy^2 = dx^2$ vanishes on the unit circle.  This is an instance where it is advantageous to rescale a~metric.}.  The power series for these analytic trigonometric functions are as follows:
\begin{gather*}
\Ck(\phi) = 1 - \frac{1}{2!}\kappa \phi^2 + \frac{1}{4!} \kappa^2 \phi^4 + \cdots,
\\
\Sk(\phi) = \phi - \frac{1}{3!}\kappa \phi^3 + \frac{1}{5!} \kappa^2 \phi^5 + \cdots.
\end{gather*}
Note that $\Ck^2(\phi) + \kappa \Sk^2(\phi) = 1$.  We also have that
\begin{gather*}
 \frac{d}{d \phi} C_{\kappa} (\phi) = -\kappa S_{\kappa}(\phi),\qquad
 \frac{d}{d \phi} S_{\kappa} (\phi) = C_{\kappa}(\phi).
\end{gather*}

\begin{definition}
Let $U_{\kappa}(1)$ denote the group (under multiplication) of unit complex numbers in~$\C_{\kappa}$.
\end{definition}

Before we proceed it might be insightful to see how the algebraic structure of $\C_{\kb}$ is useful in describing kinematics for the kinematical groups $M$ (f\/lat Minkowski spacetimes) and $G$ (f\/lat Galilean spacetime).  We will demonstrate the well known fact that classical kinematics is a~limiting case of relativistic kinematics.  The Lorentz transformation
\[
\left(
  \begin{matrix}
     x^{\prime} \\
     t^{\prime} \\
   \end{matrix}
\right) =
\frac{1}{\sqrt{1 - v^2/c^2}}
\left(
  \begin{matrix}
    1 & -v \\
    -\frac{v}{c^2} & 1 \\
  \end{matrix}
\right)
\left(
  \begin{matrix}
    x \\
    t \\
  \end{matrix}
\right)
\]
can be simply written in complex notation as $z^{\prime} = e^{i\theta}z$ in $C_{\kb}$, where $z^{\prime} = t^{\prime} + ix^{\prime}$, $z = t + ix$, and $\kb = -\frac{1}{c^2}$.  This is because rotation about the origin through the angle $\theta$ in the complex plane $\C_{\kb}$ can be written as the linear transformation (or boost)
\[
z \mapsto e^{i\theta}z \rightsquigarrow
\left(
  \begin{matrix}
     x \\
     t \\
   \end{matrix}
\right) \mapsto
\left(
  \begin{matrix}
    C_{\kb} (\theta) & S_{\kb} (\theta) \\
    -\kb S_{\kb} (\theta) & C_{\kb} (\theta) \\
  \end{matrix}
\right)
\left(
  \begin{matrix}
     x \\
     t \\
   \end{matrix}
\right)
\]
in $\R^2$, where $\theta = T_{\kb}^{-1} \left( -v \right)$, so that
\begin{gather*}
T_{\kb} (\theta) = \frac{S_{\kb} (\theta)}{C_{\kb} (\theta)} = -v \qquad \mbox{and}\\
C^2_{\kb} (\theta) + \kb S^2_{\kb} (\theta) = C^2_{\kb} (\theta) - \frac{1}{c^2} S^2_{\kb} (\theta) = 1.
\end{gather*}

On the other hand the Galilean transformation
\[
\left(
  \begin{matrix}
     x^{\prime} \\
     y^{\prime} \\
   \end{matrix}
\right) =
\left(
  \begin{matrix}
    1 & -v \\
    0 & 1 \\
  \end{matrix}
\right)
\left(
  \begin{matrix}
    x \\
    t \\
  \end{matrix}
\right)
\]
can be simply written in complex notation as $z^{\prime} = e^{i\theta}z$ in $C_0$, where $\kb = -\frac{1}{c^2} = 0$ as the speed of light is inf\/inite in Galilean spacetime.  Note that $C_0 (\theta) = 1$ and $S_0 (\theta) = \theta$, so that rotation about the origin through the angle $\theta$ can be written as the linear ``shift" transformation (or boost)
\[
z \mapsto e^{i\theta}z \rightsquigarrow
\left(
  \begin{matrix}
     x \\
     t \\
   \end{matrix}
\right) \mapsto
\left(
  \begin{matrix}
    C_0 (\theta) & S_0 (\theta) \\
    0 & C_0 (\theta) \\
  \end{matrix}
\right)
\left(
  \begin{matrix}
     x \\
     t \\
   \end{matrix}
\right) =
\left(
  \begin{matrix}
    1 & -\theta \\
    0 & 1 \\
  \end{matrix}
\right)
\left(
  \begin{matrix}
     x \\
     t \\
   \end{matrix}
\right)
\]
in $\R^2$, where $\theta = -v$.

The speed-space contraction $\kb \rightarrow 0$ (passing from relative- to absolute-time) takes Minkowski spacetime to Galilean spacetime, as $C_{\kb} (\theta) \rightarrow 1$ and $S_{\kb} (\theta) \rightarrow \theta$.  The unit circle $z\overline{z} = 1$ in Minkowski spacetime, a hyperbola, transforms into the ``degenerate'' hyperbola given by $t = \pm 1$.  The light cone in Minkowski spacetime, given by $t = \pm\frac{1}{c}x$, transforms into the ``light cone'' $t = 0$ in Galilean spacetime, where $c = \infty$.  Locally all spacetimes are equivalent to Minkowski or Galilean spacetime via space-time contractions where $\ka \rightarrow 0$ (passing from a~cosmological to a local group) so we see that $\kb = -\frac{1}{c^2}$.

Although we will not need the theorem stated below, the reader might be interested in seeing it.  Undoubtedly this theorem was known to Yaglom.

\begin{theorem} [Yaglom]
Let $f(t,x) = u(t,x) + iv(t,x)$ and $i^2 = -\kappa$, where the partial derivatives of $u$ and $v$ are continuous on an open set.  Then $f$ is holomorphic on that open set if and only if the Cauchy--Riemann equations
\begin{gather*}
u_t   = v_x, \qquad
u_x   = -\kappa v_t
\end{gather*}
are satisfied.  Furthermore, $f$ is conformal at any point $w$ where $f^{\prime}(w)$ is not a zero-divisor\footnote{On the complex plane $\C_{\kappa}$, the argument function is only def\/ined on the set of non-zero divisors, for a non-zero divisor $w$ can be written $w = re^{i\theta}$ where $r$ is the norm $\mbox{sgn}(w \overline{w})\sqrt{| w \overline{w} |^2}$ of $w$ and $\mbox{Arg}(re^{i\theta}) = \theta$,  see~\cite{HH04}.}.
\end{theorem}

The usual proofs for $\kappa = 1$ apply.


\section{A very brief review of some work\\ by Ballesteros, Herranz, Ortega and Santander}

It is the purpose of this section to introduce some material by Ballesteros, Herranz, Ortega and Santander that we will refer to in subsequent sections.  A real matrix representation for a~kinematical Lie algebra, denoted by $so_{\ka, \kb}(3)$, is given by
\[
   H =
   \left(
   \begin{matrix}
   0 & -\ka & 0 \\
   1 &     0 & 0 \\
   0 &    0 & 0
   \end{matrix}
   \right), \qquad
      P =
      \left(
      \begin{matrix}
      0 & 0 & -\ka \kb \\
      0 & 0 & 0 \\
      1 & 0 & 0
      \end{matrix}
      \right),  \qquad \mbox{and} \qquad
         K =
         \left(
         \begin{matrix}
         0 & 0 & 0 \\
         0 & 0 & -\kb \\
         0 & 1 & 0
         \end{matrix}
         \right),
 \]
where the structure constants are given by the commutators \eqref{eq1}
\[ \left[ K, H \right] = P,  \qquad \left[ K, P \right] = -\kb H, \qquad \mbox{and}  \qquad \left[ H, P \right] = \ka K.  \]

Elements of a corresponding kinematical Lie group, denoted by $SO_{\ka, \kb}(3)$, are given by real-linear, orientation-preserving isometries of $\R^3 = \{ (y, t, x)) \}$ imbued with the (possibly indef\/inite or degenerate) metric $ds^2 = dy^2 + \ka dt^2 +  \ka \kb dx^2$.
The one-parameter subgroups $\mathcal{H}$, $\mathcal{P}$, and $\mathcal{K}$ generated respectively by $H$, $P$, and $K$ consist of matrices of the form
\begin{gather*}
   e^{\alpha H} =
   \left(
   \begin{matrix}
   C_{\ka}(\alpha) & -\ka S_{\ka}(\alpha) & 0 \\
   S_{\ka}(\alpha) & C_{\ka}(\alpha)        & 0 \\
   0                       &  0                             & 1
   \end{matrix}
   \right),
\qquad
   e^{\beta P} =
   \left(
   \begin{matrix}
   C_{\ka \kb}(\beta) & 0 & -\ka \kb S_{\ka \kb}(\beta) \\
   0                           & 1 & 0        \\
   S_{\ka \kb}(\beta) & 0 & C_{\ka \kb}(\beta)
   \end{matrix}
   \right),
\end{gather*}
and
\[
   e^{\theta K} =
   \left(
   \begin{matrix}
   1 & 0 & 0 \\
   0 & C_{\kb}(\theta)  & -\kb S_{\kb}(\theta)  \\
   0 & S_{\kb}(\theta)  & C_{\kb}(\theta)
   \end{matrix}
   \right).
\]

Ballesteros, Herranz, Ortega and Santander have constructed spacetimes as homogeneous spa\-ces\footnote{See \cite{BH06, HS98, HS02}, and also \cite{HOS00}, where a~special case of the group law is investigated, leading to a plethora of trigonometric identities.} by looking at real representations of their motion groups $\SO$.  The spaces $SO_{\ka, \kb}(3) / \mathcal{K}$, $SO_{\ka, \kb}(3) / \mathcal{H}$, and $SO_{\ka, \kb}(3) / \mathcal{P}$ are homogeneous spaces for $SO_{\ka, \kb}(3)$.  When $SO_{\ka, \kb}(3)$ is a~kinematical group, then $SO_{\ka, \kb}(3) / \mathcal{K}$ can be identif\/ied with the manifold of space-time translations, $SO_{\ka, \kb}(3) / \mathcal{P}$ the manifold of time-like geodesics, and $SO_{\ka, \kb}(3) / \mathcal{H}$ the manifold of space-like geodesics.

\section{Generalized quaternions}

The goal of this section is to develop a simple algebraic description of the kinematical algebras, using what we already know about the generalized complex numbers.  To that end, we begin by putting the Hermitian norm $dz d\overline{z} = dz_1 d\overline{z_1} + \ka dz_2 d\overline{z_2}$ on $\C_{\kb}^2 \equiv \CB \times  \CB$, where $z = \left( z_1, z_2 \right)$ is an element of $\C_{\kb}^2$.  The construction below follows a natural course based on the double covering of $SO(3)$ by $SU(2)$ as part of the geometry of the standard quaternions.

The Hermitian inner product is obtained as follows.  Let $z = \left( z_1, z_2 \right)$ and $w = \left( w_1, w_2 \right)$.  Then
\begin{gather*}
\langle z, w \rangle  = \frac{1}{2} \big( \left| z + w \right|^2 - \left| z \right|^2 - \left| w \right|^2 \big) \\
\phantom{\langle z, w \rangle}{}  =
\frac{1}{2} \left( \left( z_1 + w_1 \right) \left( \overline{z_1} + \overline{w_1} \right) + \ka \left( z_2 + w_2 \right) \left( \overline{z_2} + \overline{w_2} \right) - z_1 \overline{z_1} - \ka z_2 \overline{z_2} - w_1\overline{w_1}\! - \ka w_2 \overline{w_2} \right)\!\! \\
\phantom{\langle z, w \rangle}{} = \frac{1}{2} \left( z_1 \overline{w_1} + \overline{z_1} w_1 + \ka z_2 \overline{w_2} + \ka \overline{z_2} w_2 \right) = x_1 x_2 + \kb y_1 y_2 + \ka u_1 u_2 + \ka\kb v_1 v_2,
\end{gather*}
where $z_1 = x_1 + i y_1$, $z_2 = u_1 + iv_1$, $w_1 = x_2 + iy_2$, and $w_2 = u_2 + iv_2$.  So in $\R^4$ we can write the inner product as
\[
\left(
\begin{matrix}
x_1 & y_1 & u_1 & v_1
\end{matrix}
\right)
\left(
\begin{matrix}
1 & 0 & 0 & 0 \\
0 & \kb & 0 & 0 \\
0 & 0 & \ka & 0 \\
0 & 0 & 0 & \ka\kb
\end{matrix}
\right)
\left(
\begin{matrix}
x_2 \\
y_2 \\
u_2 \\
v_2 \\
\end{matrix}
\right)
= x_1 x_2 + \kb y_1 y_2 + \ka u_1 u_2 + \ka\kb v_1 v_2.
\]

\begin{definition}
By the set of generalized quaternions $\Q$ (or simply quaternions for short) we will mean the set of numbers of the form $\{ ( x + \I y + \J u + \K v ) \; | \; \I^2 = -\kb, \J^2 = -\ka, \K^2 = -\ka\kb \}$ with the following product rules\footnote{See \cite{V93} for another description of the generalized quaternions.}
\begin{alignat*}{3}
& \I\J   = \K ,\qquad     &&  \J\I   = -\K , & \\
& \J\K  = \ka \I,\qquad  && \K\J = -\ka \I, & \\
& \K\I  = \kb \J,\qquad && \I\K   = -\kb \J.&
\end{alignat*}
We will show below that $\Q$ is a real associative algebra over the reals and that the pure quaternions represent the kinematical algebras given by equation~\eqref{eq1}.
\end{definition}

If $q = x + \I y + \J u + \K v$, then $q \overline{q} = x^2 + \kb y^2 + \ka u^2 + \ka\kb v^2$, where $\overline{q} = x - \I y - \J u - \K v$.  So if we identify points of $\Q$ with points of $\C_{\kb}^2 = \{ (z_1, z_2) \}$ by the correspondence
\[
x + \I y + \J u + \K v = z_1 + z_2 \J \rightsquigarrow \left( z_1, z_2 \right) ,
\]
where $z_1= x + i y$ and $z_2 = u + i v$ (in terms of quaternions
we can think of $z_1$ and $z_2$ as $z_1 = x + \I y$ and $z_2 = u + \I v$), then the norm of $q$ corresponds to the norm of $(z_1,z_2)$.

\begin{definition}
Let $SU_{\ka, \kb}(2)$ denote the group of all matrices of the form
\[
\left(
   \begin{matrix}
      z_1                 & z_2 \\
      -\ka \overline{z_2} & \overline{z_1}
   \end{matrix}
\right)
\]
with determinant $z_1 \overline{z_1} + \ka z_2 \overline{z_2} = 1$.   It was shown in \cite{M07} that $SU_{\ka, \kb}(2)$ is a double cover of $SO_{\ka, \kb}(3)$, and that $su_{\ka, \kb}(2)$ consists of those elements $B$ of $M(2,\C_{\kb})$ such that \mbox{$B^{\star}A {+} AB {=} 0$} where $A$ is the matrix
\[
A =
\left(
   \begin{matrix}
   \ka & 0 \\
   0    & 1
   \end{matrix}
\right).
\]
We will see below that $su_{\ka, \kb}(2)$ can be identif\/ied with the space of pure quaternions, a real algebra, and that f\/inally the space of pure quaternions is a kinematical algebra.
\end{definition}

Under the correspondence
\[
x + \I y + \J u + \K v \rightsquigarrow
\left(
   \begin{matrix}
      z_1                 & z_2 \\
      -\ka \overline{z_2} & \overline{z_1}
   \end{matrix}
\right)
\]
the set of unit quaternions is identif\/ied with $SU_{\ka, \kb}(2)$.  The context should make it clear as to whether elements of $SU_{\ka, \kb}(2)$ are to be treated as elements of $M(2, \C_{\kb})$ or as unit quaternions in $\Q$.  The inner product on $\C_{\kb}^2$ corresponds in $\Q$ to
\begin{gather*}
\langle q_1, q_2 \rangle  =  \frac{1}{2} \left( q_1 \overline{q_2} + q_2\overline{q_1} \right)
  = \frac{1}{2} \left( \left(z_1 + z_2 \J \right) \left( \overline{w_1} - w_2 \J \right) + \left( w_1 + w_2 \J \right)\left(\overline{z_1} - z_2 \J \right)  \right) \\
\phantom{\langle q_1, q_2 \rangle}{}  =  \frac{1}{2} \left( z_1 \overline{w_1} + \overline{z_1} w_1 + \ka z_2 \overline{w_2} + \ka \overline{z_2} w_2 \right)
   = \frac{1}{2} \big( \left| z + w \right|^2 - \left| z \right|^2 - \left| w \right|^2 \big)
   = \langle z, w \rangle,
\end{gather*}
since $\J \left( x + \I y \right) = \left( x - \I y \right) \J$ and $\J^2 = -\ka$.

We can see that $\Q$ and the subspace of $M(2, \Cb)$ consisting of all matrices of the form
$
\left(
   \begin{matrix}
      z_1                 & z_2 \\
      -\ka \overline{z_2} & \overline{z_1}
   \end{matrix}
\right)
$
are isomorphic as algebras, for if $q_1 = z_1 + z_2 \J$ and $q_2 = w_1 + w_2 \J$ are two quaternions with corresponding matrices
\[
\left(
   \begin{matrix}
      z_1                 & z_2 \\
      -\ka \overline{z_2} & \overline{z_1}
   \end{matrix}
\right)
\qquad \mbox{and} \qquad
\left(
   \begin{matrix}
      w_1                 & w_2 \\
      -\ka \overline{w_2} & \overline{w_1}
   \end{matrix}
\right),
\]
then
\[ q_1 + q_2 \rightsquigarrow
\left(
   \begin{matrix}
      z_1                 & z_2 \\
      -\ka \overline{z_2} & \overline{z_1}
   \end{matrix}
\right)
+
\left(
   \begin{matrix}
      w_1                 & w_2 \\
      -\ka \overline{w_2} & \overline{w_1}
   \end{matrix}
\right)
\]
and
\[ q_1 q_2 \rightsquigarrow
\left(
   \begin{matrix}
      z_1                 & z_2 \\
      -\ka \overline{z_2} & \overline{z_1}
   \end{matrix}
\right)
\left(
   \begin{matrix}
      w_1                 & w_2 \\
      -\ka \overline{w_2} & \overline{w_1}
   \end{matrix}
\right)
=
\left(
   \begin{matrix}
      z_1w_1 - \ka z_2 \overline{w_2}                 & z_1 w_2 + z_2 \overline{w_1} \\
      -\ka \overline{z_2}w_1 - \ka \overline{z_1} \overline{w_2} & \overline{z_1} \overline{w_1} - \ka \overline{z_2} w_2
   \end{matrix}
\right), \]
since
\[ \left( z_1 + z_2 \J \right) \left( w_1 + w_2 \J \right) = \left( z_1 w_1 - \ka z_2 \overline{w_2} \right) + \left( z_1 w_2 + z_2 \overline{w_1} \right) \J.
 \]

\begin{definition}  We def\/ine the unit one-sphere and two-sphere\footnote{The context should make it clear as to whether these spheres are to thought of in terms of generalized complex or quaternion numbers.}
\begin{gather*}
 S^1_{\kb} = U_{\kb}(1) = \{ z \in \C_{\kb} , \; \left| z \right| = 1 \} \rightsquigarrow \{ e^{\I \theta} \} \subset \Q,\\
 S^3_{\ka, \kb} =
\{ (z, w) \in \C^2_{\kb} , \; |(z,w)| = 1 \}  \rightsquigarrow \{ q \in \Q \; | \; \left| q \right| = 1 \},
\end{gather*}
where the set of unit quaternions is given by numbers of the form $e^{\I y + \J u + \K v}$.  So $S^3_{\ka, \kb}$ can be identif\/ied with $SU_{\ka, \kb}(2)$.
\end{definition}

The plane spanned by $1$ and $\I$ can be easily identif\/ied with $\C_{\kb}$, and the intersection of this plane with the sphere of unit quaternions then corresponds to the unit circle of $\C_{\kb}$.  Similar remarks hold for $\C_{\ka}$ or $\C_{\ka \kb}$ for the planes spanned by 1 and $\J$ or 1 and $\K$ respectively.  So $e^{\I t}$, $e^{\J t}$, and $e^{\K t}$ are all unit quaternions.

If $a$ is a unit quaternion, then $a^{-1} = \overline{a}$.  Since $\overline{ab} = \overline{b}\overline{a}$ for any two quaternions $a$ and $b$, it follows that $| a q b^{-1} | = | q |$ for any quaternion $q$, provided that both $a$ and $b$ are unit quaternions.  In fact, the generalized quaternions are a composition algebra, so that $\left| q_1 q_2 \right| = \left| q_1 \right| \left| q_2 \right|$.  Also, $a \mathbf{q} a^{-1}$ is a pure quaternion since $\overline{a\mathbf{q}a^{-1}} = -a\mathbf{q}a^{-1}$, where $\mathbf{q}$ denotes the pure part of $q$.  So the linear transformations (in terms of real coordinates)
\[
\R^4 \rightarrow \R^4 \;\; \mbox{def\/ined by the automorphism} \;\; q \rightarrow a q b^{-1}
\]
and
\[
\R^3 \rightarrow \R^3 \;\; \mbox{def\/ined by the inner automorphism} \;\; {\bf q} \rightarrow a {\bf q} a^{-1}
\]
respectively give rotations of $\Q$ and the subspace of pure quaternions.  It might appear then that $SU_{\ka, \kb}(2) \times SU_{\ka, \kb}(2)$ is a double cover of the group of rotations of $\C_{\kb}^2$ with Hermitian metric $dz_1 d\overline{z_1} + \ka dz_2 d\overline{z_2}$, as $(a,b)$ and $(-a,-b)$ represent the same rotation, but not all rotations can be so represented by such an automorphism.  For example, if both $\ka$ and $\kb$ vanish, then rotations of $\R^4$ are of the form
\[
\left(
   \begin{matrix}
   1 & 0          & 0           & 0          \\
   0 & m_{22} & m_{23} & m_{24} \\
   0 & m_{32} & m_{33} & m_{34} \\
   0 & m_{42} & m_{43} & m_{44}
   \end{matrix}
   \right),
\]
and so the rotation group is $9$-dimensional.  Yet $SU_{\ka, \kb}(2) \times SU_{\ka, \kb}(2)$ has dimension $6$.  Similarly $SU_{\ka, \kb}(2)$ is not a double cover for the rotation group for the subspace of pure quaternions\footnote{Note that ${\bf q} \overline{\bf q} = \kb y^2 + \ka u^2 + \ka\kb w^2$ so that, when both $\ka$ and $\kb$ vanish, the dimension of the rotation group of the pure quaternions is $9$-dimensional, as any orientation preserving linear map of $\R^3$ is a rotation.}.

Let $su_{\ka, \kb}(2)$ denote the Lie algebra of $SU_{\ka, \kb}(2)$.  If we identify $SU_{\ka, \kb}(2)$ with $S^3_{\ka, \kb}$, the space of unit quaternions, then $su_{\ka, \kb}(2)$ can be represented by the space of pure quaternions:  For if $\mathbf{q}$ is a pure quaternion, then $e^{\mathbf{q}}$ is a unit quaternion, as $\mathbf{q} \overline{\mathbf{q}} = \overline{\mathbf{q}} \mathbf{q}$ so that $e^{\mathbf{q}} \overline{e^{\mathbf{q}}} = e^{\mathbf{q}} e^{\overline{\mathbf{q}}} = e^{\mathbf{q} + \overline{\mathbf{q}}} = e^{\mathbf{0}} = 1$.  $SU_{\ka, \kb}(2)$ acts on its Lie algebra $su_{\ka, \kb}(2)$ by the inner automorphism $\mathbf{p} \mapsto e^{\frac{\theta}{2}\mathbf{q}} \mathbf{p} e^{-\frac{\theta}{2}\mathbf{q}}$ where both $\mathbf{p}$ and $\mathbf{q}$ are pure quaternions.  Since
\[
\left. \frac{d}{d\theta} \right|_{\theta = 0} e^{\frac{\theta}{2}\mathbf{q}} \mathbf{p} e^{-\frac{\theta}{2}\mathbf{q}} = \frac{1}{2} \left( \mathbf{q}\mathbf{p} - \mathbf{p}\mathbf{q} \right) = \frac{1}{2} \left[ \mathbf{q}, \mathbf{p} \right],
\]
then
\begin{gather*}
\frac{1}{2} \left[ \I , \J \right]  = \left. \frac{d}{d\theta} \right|_{\theta = 0} e^{\frac{\theta}{2} \I} \J e^{-\frac{\theta}{2}\I}  = \left. \frac{d}{d\theta} \right|_{\theta = 0} e^{\frac{\theta}{2} \I}  e^{\frac{\theta}{2}\I} \J  = \I \J  = \K, \\
\frac{1}{2} \left[ \I , \K \right]  = \left. \frac{d}{d\theta} \right|_{\theta = 0} e^{\frac{\theta}{2} \I} \K e^{-\frac{\theta}{2}\I}  = \left. \frac{d}{d\theta} \right|_{\theta = 0} e^{\frac{\theta}{2} \I}  e^{\frac{\theta}{2}\I} \K  = \I \K   = -\kb \J, \\
\frac{1}{2} \left[ \J , \K \right]  = \left. \frac{d}{d\theta} \right|_{\theta = 0} e^{\frac{\theta}{2} \J} \K e^{-\frac{\theta}{2}\J}  = \left. \frac{d}{d\theta} \right|_{\theta = 0} e^{\frac{\theta}{2} \J}  e^{\frac{\theta}{2}\J} \K  = \J \K  = \ka \I
\end{gather*}
as $\J z = \overline{z} \J$ and $\K z = \overline{z} \K$.  We can then represent a given kinematical Lie algebra by
\[
 K \rightsquigarrow 2\I , \qquad H \rightsquigarrow 2\J , \qquad \mbox{and} \qquad P \rightsquigarrow 2\K.
\]

In terms of the ordered basis $\{ E_1, E_2, E_3 \} = \{ 2\I, 2\J, 2\K \}$ for $su_{\ka, \kb}(2)$, the structure constants are given by $\left[ E_i, E_j \right] = C_{ij}^kE_k$, and so the Killing form on $su_{\ka, \kb}(2)$ is given by $g_{ij} = \sum_{r,s} C_{is}^r C_{jr}^s$~or
\[
\left( g_{ij} \right) = -2
\left(
\begin{matrix}
\kb & 0 & 0 \\
0 & \ka & 0 \\
0 & 0 & \ka\kb
\end{matrix}
\right).
\]
The Killing form is preserved by the inner automorphism.

We may form three natural homogeneous spaces:
\begin{gather*}
 SU_{\ka, \kb}(2)/\langle \I \rangle  =  SU_{\ka, \kb}(2)/\mathcal{K}  \rightsquigarrow  S^3_{\ka, \kb}/S^1_{\kb},\\
 SU_{\ka, \kb}(2)/\langle \J \rangle  =  SU_{\ka, \kb}(2)/\mathcal{H}  \rightsquigarrow  S^3_{\ka, \kb}/S^1_{\ka},\\
 SU_{\ka, \kb}(2)/ \langle \K \rangle  =  SU_{\ka, \kb}(2)/\mathcal{P}  \rightsquigarrow  S^3_{\ka, \kb}/S^1_{\ka\kb} .
\end{gather*}
The Killing form naturally determines a metric for each of these homogeneous spaces with respective inner products
\[
\left(
\begin{matrix}
\ka & 0  \\
0 & \ka\kb  \\
\end{matrix}
\right),
\qquad
\left(
\begin{matrix}
\kb & 0  \\
0 & \ka\kb  \\
\end{matrix}
\right)
\qquad \mbox{and} \qquad
\left(
\begin{matrix}
\kb & 0  \\
0 & \ka  \\
\end{matrix}
\right).
\]
Following Ballesteros, Herranz, Ortega, and Santander we will rescale (even if $\ka$ or $\kb$ is equal to zero) so that, in fact, the respective inner products are as follows:
\[
\left(
\begin{matrix}
1 & 0  \\
0 & \kb  \\
\end{matrix}
\right),
\qquad
\left(
\begin{matrix}
1 & 0  \\
0 & \ka \\
\end{matrix}
\right)
\qquad \mbox{and} \qquad
\left(
\begin{matrix}
\kb & 0  \\
0 & \ka  \\
\end{matrix}
\right).
\]
The resulting metrics can be indef\/inite as well as degenerate.

\begin{theorem}
Let $H$, $P$, and $K$ denote the respective generators for time translations, space translations, and boosts of the kinematical algebra with commutators
\[ \left[ K, H \right] = P,  \qquad \left[ K, P \right] = -\kb H, \qquad \mbox{and}  \qquad \left[ H, P \right] = \ka K. \]
Then the kinematical algebra can be represented as the space of pure quaternions in $\Q$ by
\[
 K \rightsquigarrow 2\I , \qquad H \rightsquigarrow 2\J , \qquad \mbox{and} \qquad P \rightsquigarrow 2\K.
\]
If $SU_{\ka, \kb}(2)$ denotes the group of unit quaternions with lie algebra $su_{\ka, \kb}(2)$, then $su_{\ka, \kb}(2)$ is the space of pure quaternions and the homogeneous space $SU_{\ka, \kb}/\langle \I \rangle$ is the space of events, $SU_{\ka, \kb}/\langle \J \rangle$ is the space of space-like geodesics, and $SU_{\ka, \kb}/\langle \K \rangle$ is the space of time-like geo\-de\-sics.
\end{theorem}

\section{The generalized Clif\/ford f\/ibration}

As pointed out by Urbantke (see~\cite{U03}), Penrose has called the Clif\/ford f\/ibration an ``element of the architecture of our world''.  This f\/ibration can be used to describe two-level quantum systems, the harmonic oscillator, Taub-NUT space, Robinson congruences, helicity representations, magnetic monopoles, and the Dirac equation.  By generalizing the Clif\/ford f\/ibration we will give yet another physical application by modeling all kinematical algebras save for the static algebra.

\begin{figure}[t]
\centering
\includegraphics[scale=0.80]{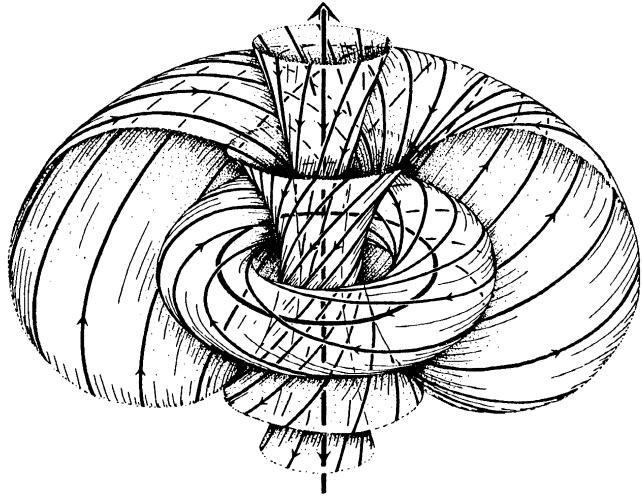}

\caption{Stereographic projection of the Clif\/ford f\/ibration for $S^3$ onto $\R^3$.  Image courtesy of [Penrose~R., Rindler~W.,  Spinors and space-time, Vol.~2, Spinor and twistor methods in space-time geometry,   Cambridge University Press, 1986].}
\end{figure}

$S_{\kb}^1$ acts freely and smoothly on $S_{\ka, \kb}^3$ by left multiplication: $q = z + w \J \mapsto e^{\I \theta}z + e^{\I \theta} w \J$.  If $e^{\I \theta} q = q$, then $e^{\I \theta} = 1$ since $\left| q \right| = 1$ and so $q$ is not a zero-divisor.  So $S_{\ka, \kb}^3$ is the total space of a principal $S_{\kb}^1$ bundle.  So what can we say about the base space of this bundle?

We def\/ine $\Cb\mathbb{P}_1$ as the space of all complex one-dimensional subspaces of the vector spa\-ce~$\C_{\kb}^2$.  Each subspace is uniquely described as the solution space to the complex linear equation $Az_1 + Bz_2 = 0$ where $\left[ \frac{A}{B} \right]$ def\/ines a point on the Riemann sphere~$\Sigma_{\kb}$.  Note that $A$ and $B$ cannot both be zero-divisors, for then $\left[ \frac{A}{B} \right]$ is not def\/ined, and the set of points $(z_1, z_2)$ satisfying $Az_1 + Bz_2 = 0$ is no longer one-dimensional.

Alternatively, if we think of the complex line through the point $(z_1, z_2)$ as being given by points of the form $\lambda(z_1, z_2)$, where $\lambda$ takes all values in $\Cb$, then we get a line precisely when either $z_1$ or $z_2$ is not a zero-divisor.    For such a line we may let $A = z_2/z_1$ and $B = -1$ if $z_1$ is not a zero-divisor, and $A = -1$ and $B = z_1/z_2$ otherwise.  Taking values in the Riemann sphere (so that ``inf\/inities" are allowed) we may always write $\left[ \frac{A}{B} \right] = -\left[ \frac{z_2}{z_1} \right]$.  Note that distinct lines always intersect at the origin, but they may also intersect at points where both coordinates are zero-divisors, for $\lambda$ is allowed be a zero-divisor:  So two points do not necessarily determine a unique line.

 So we may identify $\Cb\mathbb{P}_1$ with the Riemann sphere.  A complex line will intersect $S_{\ka, \kb}^3$ exactly when $|z_1|^2 + \ka |z_2|^2 = 1$ for some point $( z_1, z_2)$ on the line:  Let us call a line {\it null} if $|z_1|^2 + \ka |z_2|^2 = 0$ for all points $(z_1, z_2)$ on that line.  So only the null lines do not intersect the unit sphere.  A complex line that does intersect the unit sphere does so at points of the form $\{ \left( e^{i \theta}z_1, e^{i \theta}z_2 \right) \}$, where $(z_1, z_2)$ is any point belonging to the intersection.  Let us denote by $\Sigma_{\ka, \kb}$ the subset of $\Sigma_{\kb}$ corresponding to non-null lines.  We will see below that $\Sigma_{\ka, \kb}$ is the homogeneous space $S^3_{\ka, \kb}/S^1_{\kb}$.

So given a null line we must have that both $Az_1 + Bz_2 = 0$ and $\left| z_1 \right|^2 + \ka \left| z_2 \right|^2 = 0$ for all points $\left( z_1, z_2 \right)$ on the line.  Recall that $w \equiv \left[ \frac{A}{B} \right] = -\left[ \frac{z_2}{z_1} \right]$ represents a point on the Riemann sphere $\Sigma_k$.  When $\ka = 0$, $\Sigma_{0, \kb}$ is $\Sigma_{\kb}$ with all inf\/inities removed.  If $\ka \neq 0$, then null lines can exist only when $\ka < 0$, and in this case $\Sigma_{\ka, \kb}$ is $\Sigma_{\kb}$ with all points of the form $|w|^2 = -\frac{1}{\ka}$ removed.  It might be useful at this point to consider a familiar case:  When $\kb = 1$ and $\ka = 1$, $0$, or $-1$ we have elliptic, euclidian, and hyperbolic geometry respectively.   For elliptic geometry $\Sigma_{1,1} = \Sigma_1$ is the well known Riemann sphere.  For the euclidean plane $\Sigma_{0,1} = \Sigma_1 \setminus \{ \infty \}$ is topologically a plane.  And for the hyperbolic plane $\Sigma_{-1, 1} = \Sigma_1 \setminus S^1$ is topologically a union of two planes (each plane giving rise to a model of the hyperbolic plane).

We observe that the vectors $\I q$, $\J q$, and $\K q$ span the tangent space $T_q(S^3_{\ka, \kb})$ of $S^3_{\ka, \kb}$ at $q$.  For if $q \in S^3_{\ka, \kb}$, then $e^{\I t}q$, $e^{\J t}q$, and $e^{\K t}q$ are unit quaternions.  Keeping $q$ f\/ixed and letting $t$ vary, the respective tangents to the curves $e^{\I t}q$, $e^{\J t}q$, and $e^{\K t}q$ passing through $q$ are given by $\left. \frac{d}{dt} \right|_{t=0} e^{\I t}q = \I q$, $\left. \frac{d}{dt} \right|_{t=0} e^{\J t}q = \J q$, and $\left. \frac{d}{dt} \right|_{t=0} e^{\K t}q = \K q$.  Now
\begin{gather*}
\I q  =   \I \left( x + \I y + \J u + \K v \right)  =  \I x - \kb y + \K u -\kb \J v  =   z_1 \I+  z_2 \K, \\
\J q  =   \J \left( x + \I y + \J u + \K v \right)  =  \J x - \K y - \ka u + \ka \I v  =  -\ka \overline{z_2} + \overline{z_1} \J, \\
\K q  =  \K \left( x + \I y + \J u + \K v \right)  =  \K x + \J \kb y - \I \ka u - \ka\kb v  =  -\ka  \overline{z_2} \I +  \overline{z_1} \K,
\end{gather*}
and so $\I q$, $\J q$, and $\K q$ are linearly independent since $| q | = 1$.  Note that $\I q$, $\J q$ and $\K q$ are mutually orthogonal since
\begin{gather*}
\langle \I q, \J q \rangle   = \I q \left( -\overline{q} \J \right) + \J q \left( -\overline{q} \I \right)  = - \I \J - \J \I  = 0, \\
\langle \I q, \K q \rangle  = \I q \left( -\overline{q} \K \right) + \K q \left( -\overline{q} \I \right)    = -\I \K - \K \I   = 0, \\
\langle \J q, \K q \rangle   = \J q \left( -\overline{q} \K \right) + \K q \left( -\overline{q} \J \right)   = - \J \K - \K \J   = 0.
\end{gather*}
However, the frame $\{ \I q, \J q, \K q \}$ is not orthonormal since $| \I q |^2 = \kb$, $| \J q |^2 = \ka$, and $| \K q |^2 = \ka\kb$ (compare with the Killing form on $su_{\ka, \kb}(2)$).  The tangent plane spanned by $\J q$ and $\K q$ has the complex structure $\Cb$ since multiplying $\J q$ on the left by $\I$ yields $\K q$.  We also see that $S^3_{\ka, \kb}$ is parallelizable:  This is no surprise however, since $S^3_{\ka, \kb}$ is topologically either $S^3$, $\R^3$, or $S^1 \times \R^2$.

\begin{definition}
The Clif\/ford f\/ibration is given by
\[
\begin{CD}
S^3_{\ka, \kb} \\
\pi@VVV  \\
\Sigma_{\ka, \kb}
\end{CD}
\;\;\;\;
\rightsquigarrow
\;\;\;\;
 \begin{CD}
S^3_{\ka, \kb} \\
\pi@VVV  \\
S^3_{\ka, \kb}/S^1_{\kb}
\end{CD}
\;\;\;\;
\rightsquigarrow
\;\;\;\;
 \begin{CD}
S^3_{\ka, \kb} \\
\pi@VVV  \\
S^3_{\ka, \kb}/\langle \I \rangle
\end{CD}
\;\;\;\;
\rightsquigarrow
\;\;\;\;
 \begin{CD}
S^3_{\ka, \kb} \\
\pi@VVV  \\
S^3_{\ka, \kb}/ \mathcal{K}
\end{CD}
\]
where $\pi^{-1} \left( \left[ \frac{z_1}{z_2} \right] \right) = \{ e^{\I \phi}(z_1 + z_2\J) \}$.
\end{definition}

This is a principal f\/iber bundle over $\Sigma_{\ka, \kb}$ with f\/iber given by $S^1_{\kb}$ (the curve $e^{\I t} q$ is the f\/iber passing through $q$).  The Clif\/ford f\/low is given by the vector f\/ield $\chi_\I (q) = \I q$, and the canonical connection is determined by the horizontal planes spanned by $\J q$ and $\K q$ at each unit quaternion $q \in S^3_{\ka, \kb}$.  Each such plane has the complex structure of $\C_{\kb}$.  Here $\Sigma_{\ka, \kb}$ is the spacetime for the kinematical algebra.

Similarly, we may form the f\/ibrations
\[
 \begin{CD}
S^3_{\ka, \kb} \\
\pi@VVV  \\
S^3_{\ka, \kb}/\langle \J \rangle
\end{CD}
\qquad \mbox{and} \qquad
\begin{CD}
S^3_{\ka, \kb} \\
\pi@VVV  \\
S^3_{\ka, \kb}/\langle \K \rangle
\end{CD}
\]
with respective Clif\/ford f\/lows given by $\chi_{\J} (q) = \J q$ and $\chi_{\K} (q) = \K q$.  These f\/ibrations are principle f\/iber bundles of $S_{\ka, \kb}^3$ with respective f\/ibers $S^1_{\ka}$ and $S^1_{\ka\kb}$ and they give the space of space-like and time-like geodesics of the spacetime $\Sigma_{\ka, \kb}$, as $H \rightsquigarrow 2\J$ and $P \rightsquigarrow 2\K$.  Note however that the bases $S^3_{\ka, \kb}/ \langle \J \rangle$ and $S^3_{\ka, \kb}/ \langle \K \rangle$ are not given by $\Sigma_{\ka, \kb}$, as the f\/ibers do not lie in the complex lines $Az_1 + Bz_2 = 0$ which have the complex structure of $\Cb$, not of $\Ca$ nor $\Cab$.

\begin{theorem}
Let $H \rightsquigarrow 2\J$, $P \rightsquigarrow 2\K$, and $K \rightsquigarrow 2\I$ denote the respective generators for time translations, space translations, and boosts of the kinematical algebra with commutators
\[ \left[ K, H \right] = P,  \qquad \left[ K, P \right] = -\kb H, \qquad \mbox{and}  \qquad \left[ H, P \right] = \ka K. \]
We can construct principal fiber bundles
\[
\begin{CD}
S^3_{\ka, \kb} \\
\pi@VVV  \\
S^3_{\ka, \kb}/\langle \I \rangle
\end{CD}
\qquad \mbox{and} \qquad
\begin{CD}
S^3_{\ka, \kb} \\
\pi@VVV  \\
S^3_{\ka, \kb}/\langle \J \rangle
\end{CD}
\qquad \mbox{and} \qquad
\begin{CD}
S^3_{\ka, \kb} \\
\pi@VVV  \\
S^3_{\ka, \kb}/\langle \K \rangle
\end{CD}
\]
on the space $S^3_{\ka, \kb}$ of unit quaternions.  Here the respective base spaces are the space of events, the space of space-like geodesics, and the space of time-like geodesics with corresponding Clifford flows on $S^3_{\ka, \kb}$ given by $\chi_\I (q) = \I q$, $\chi_\J (q) = \J q$, and $\chi_\K (q) = \K q$.  The principal connections are determined by the distribution of horizontal planes spanned by $\{ \J q, \K q \}$, $\{ \I q, \K q\}$, and $\{ \I q, \J q \}$ with corresponding complex structures $\C_{\kb}$, $\C_{\ka}$, and $\C_{\ka \kb}$ for these planes.
\end{theorem}

\subsection[Optional reading on coordinate charts for $\Sigma_{\ka, \kb}$]{Optional reading on coordinate charts for $\boldsymbol{\Sigma_{\ka, \kb}}$}

Complex lines will intersect $S_{\ka, \kb}^3$ exactly when $Az_1 + Bz_2 = 0$ and $z_1\overline{z}_1 + \ka z_2\overline{z}_2 = 1$.   Let~$\omega$ denote $\frac{1}{w}$.   We can cover $\Sigma_{\ka, \kb}$ with two coordinate charts:   Let $U_1$ denote the set of points $\left[ \frac{B}{A} \right]$ of $\Sigma_{\ka, \kb}$ where $A$ is not a~zero divisor, and $U_2$ the set of points $\left[ \frac{A}{B} \right]$ where $B$ is not a~zero divisor.  Then the coordinate charts $\phi_1 : U_1 \rightarrow \Cb$ and $\phi_2 : U_2 \rightarrow \Cb$ are given by $\phi_1 \left( \left[ \frac{B}{A} \right] \right) = \omega = -\frac{z_1}{z_2} \in \C_{\kb}$ and $\phi_2 \left( \left[ \frac{A}{B} \right] \right) = w = -\frac{z_2}{z_1} \in \C_{\kb}$ respectively\footnote{Note that the map $w \mapsto \omega = \frac{1}{w} = \frac{\overline{w}}{w\overline{w}}$ is conformal on the set of non-zero divisors.}.  If $S^3_{\ka, \kb} \stackrel{\pi}{\longrightarrow} \Sigma_{\ka, \kb}$ def\/ines the projection map, then
 \[ \left( \phi_1 \circ \pi \right)^{-1}(\omega ) = \left\{ e^{\I \theta} \frac{1 + \omega \J}{\sqrt{1 + \ka |\omega|^2}} \right\} \qquad
\mbox{and} \qquad  \left( \phi_2 \circ \pi \right)^{-1}(w) = \left\{ e^{\I \theta} \frac{w + \J}{\sqrt{|w|^2 + \ka }} \right\}, \]
where $e^{\I \theta}$ is an arbitrary element of $S^1_{\kb}$.  We can then give a product structure to $\pi^{-1} \left( U_1 \right) \subset S^3_{\ka, \kb}$ and to $\pi^{-1} \left( U_2 \right) \subset S^3_{\ka, \kb}$ by
\[ \Phi_1\left( \omega, e^{\I \theta} \right)= e^{\I \theta} \frac{1 + \omega \J}{\sqrt{1 + \ka |\omega|^2}}  \qquad
\mbox{and}\qquad  \Phi_2\left( w, e^{\I \theta} \right) = e^{\I \theta} \frac{w + \J}{\sqrt{|w|^2 + \ka }}  \]
respectively.  This trivializing cover of $S^3_{\ka, \kb}$ has a gluing map
\[ \Phi_2^{-1} \circ \Phi_1 \left( \omega, e^{\I \theta} \right) = \left( \frac{1}{\omega}, e^{\I \theta} \frac{\omega}{|\omega|} \right). \]
We can also map $\Sigma_{\ka, \kb}$ to the unit sphere in $\R^3$ with metric\footnote{Recall that $SO_{\ka, \kb}(3)$ is the group of isometries of $\R^3 = \{ (y, t, x) \; | \; y, t, x \in \R \}$ with metric $dy^2 + \ka dt^2 + \ka\kb dx^2$.} $dy^2 + \ka dt^2 + \ka\kb dx^2$ by
\[ \omega = -\frac{z_1}{z_2} \mapsto \left( \mathcal{R} \frac{2\omega}{1 + \ka |\omega|^2}, \mathcal{I} \frac{2\omega}{1 + \ka |\omega|^2}, \frac{-1 + \ka |\omega|^2}{1 + \ka |\omega|^2}  \right)  \]
or
\[ w = -\frac{z_2}{z_1} \mapsto \left( \mathcal{R} \frac{2w}{1 + \ka |w|^2}, \mathcal{I} \frac{2w}{1 + \ka |w|^2}, \frac{-1 + \ka |w|^2}{1 + \ka |w|^2}  \right)  \]
as can be checked directly.

\section{The principal connection form}

The right invariant one-forms on $S^3_{\ka, \kb}$ are given by
\[
(dU) U^{-1} =
\left(
   \begin{matrix}
      dz_1                 & dz_2 \\
      -\ka d\overline{z_2} & d\overline{z_1}
   \end{matrix}
\right)
\left(
   \begin{matrix}
      \overline{z_1}                 & -z_2 \\
      \ka \overline{z_2} & z_1
   \end{matrix}
\right) =
\left(
   \begin{matrix}
      \overline{z_1} dz_1 + \ka \overline{z_2} dz_2     & -z_2 dz_1 + z_1 dz_2 \\
      -\ka \overline{z_1} d\overline{z_2} + \ka \overline{z_2} d\overline{z_1} & \ka z_2 d\overline{z_2} + z_1 d\overline{z_1}
   \end{matrix}
\right)
\]
and the left invariant one-forms on $S^3_{\ka, \kb}$ are given by
\[
U^{-1} dU =
\left(
   \begin{matrix}
      \overline{z_1}                 & -z_2 \\
      \ka \overline{z_2} & z_1
   \end{matrix}
\right)
\left(
   \begin{matrix}
      dz_1                 & dz_2 \\
      -\ka d\overline{z_2} & d\overline{z_1}
   \end{matrix}
\right)
=
\left(
   \begin{matrix}
      \overline{z_1} dz_1 + \ka z_2 \overline{dz_2}     & -z_2 d\overline{z_1} + \overline{z_1} dz_2 \\
      -\ka z_1 d\overline{z_2} + \ka \overline{z_2} dz_1 & \ka \overline{z_2} dz_2 + z_1 d\overline{z_1}
   \end{matrix}
\right).
\]

We will show that the principal connection form is given by the right invariant form $\lambda = \overline{z_1} dz_1 + \ka z_2 \overline{dz_2}$.  Let $J$ denote the almost complex structure on $\R^4$ that is compatible with multiplying $(z_1, z_2)$ by $i$ in $\C_{\kb}^2$.  Then
\begin{gather*}
i (z_1, z_2) = (-\kb y + i x, -\kb v + i u)  \\
\phantom{i (z_1, z_2) =}{}  \rightsquigarrow JX = J
   \left(
     \begin{matrix}
     x \\
     y \\
     u \\
     v
     \end{matrix}
   \right)
   =
   \left(
     \begin{BMAT}(e){cc:cc}{cc:cc}
     0 & -\kb & 0 & 0 \\
     1 & 0 & 0 & 0 \\
     0 & 0 & 0 & -\kb \\
     0 & 0 & 1 & 0
     \end{BMAT}
   \right)
   \left(
     \begin{matrix}
     x \\
     y \\
     u \\
     v
     \end{matrix}
   \right)
   =
   \left(
     \begin{matrix}
     -\kb y \\
     x \\
     - \kb v \\
     u
     \end{matrix}
   \right),
\end{gather*}
where $J^2 = -\kb I$ and $I$ is the identity matrix.  Recall that the Hermitian inner product on $\C_{\kb}^2$ is given by
\begin{gather*}
\langle z, w \rangle  = z_1 \overline{z_2} + \ka w_1 \overline{w_2} \\
\phantom{\langle z, w \rangle}{}  = \left( x_1x_2 + \kb y_1y_2 + \ka u_1u_2 + \ka\kb v_1v_2 \right)
                                 + i \left(  y_1x_2 - x_1y_2 + \ka v_1u_2 - \ka u_1v_2  \right),
\end{gather*}
where $z = (z_1, z_2) = (x_1 + i y_1, u_1 + i v_1)$ and $w = (w_1, w_2) = (x_2 + i y_2, u_2 + i v_2)$.  In real coordinates we can write this as $\langle X_1 , X_2 \rangle = \langle \langle X_1, X_2 \rangle \rangle + i \Phi (X_1, X_2)$, where $\langle \langle X_1, X_2 \rangle \rangle$ and $\Phi ( X_1, X_2)$ give the respective real and imaginary parts of the inner product $\langle X_1, X_2 \rangle$, and where
\[
X_1 = \left(
     \begin{matrix}
     x_1 \\
     y_1 \\
     u_1 \\
     v_1
     \end{matrix}
   \right),
  \qquad
X_2 = \left(
     \begin{matrix}
     x_2 \\
     y_2 \\
     u_2 \\
     v_2
     \end{matrix}
   \right).
\]
So
\[
\langle \langle X_1, X_2 \rangle \rangle = X_1^T
\left(
  \begin{matrix}
   1 & 0 & 0 & 0 \\
   0 & \kb & 0 & 0 \\
   0 & 0 & \ka & 0 \\
   0 & 0 & 0 & \ka \kb
  \end{matrix}
\right)
X_2
\]
and
\begin{gather*}
\Phi\left( X_1, X_2 \right)   = -\frac{1}{\kb} \langle \langle JX_1, X_2 \rangle \rangle
                                          = \frac{1}{\kb} \langle \langle X_1, JX_2 \rangle \rangle \\
\phantom{\Phi\left( X_1, X_2 \right)}{}  = -\frac{1}{\kb} \left( -\kb y_1 \;\; x_1 \;\; -\kb v_1 \;\; u_1 \right)
                                                \left(
                                                   \begin{matrix}
                                                       1 & 0 & 0 & 0 \\
                                                       0 & \kb & 0 & 0 \\
                                                       0 & 0 & \ka & 0 \\
                                                       0 & 0 & 0 & \ka \kb
                                                   \end{matrix}
                                                 \right)
                                                 \left(
                                                   \begin{matrix}
                                                      x_2 \\
                                                      y_2 \\
                                                      u_2 \\
                                                      v_2
                                                    \end{matrix}
                                                  \right) \\
\phantom{\Phi\left( X_1, X_2 \right)}{} = -\frac{1}{\kb} X_1^T
                                                 \left(
                                                   \begin{BMAT}(e){cc:cc}{cc:cc}
                                                      0 & 1 & 0 & 0 \\
                                                      -\kb & 0 & 0 & 0 \\
                                                      0 & 0 & 0 & 1 \\
                                                      0 & 0 &-\kb & 0
                                                   \end{BMAT}
                                                 \right)
                                               \left(
                                                   \begin{matrix}
                                                       1 & 0 & 0 & 0 \\
                                                       0 & \kb & 0 & 0 \\
                                                       0 & 0 & \ka & 0 \\
                                                       0 & 0 & 0 & \ka \kb
                                                   \end{matrix}
                                                 \right)
                                                 X_2 \\
\phantom{\Phi\left( X_1, X_2 \right)}{} = -\frac{1}{\kb} X_1^T
                                                  \left(
                                                   \begin{BMAT}(e){cc:cc}{cc:cc}
                                                      0 & \kb & 0 & 0 \\
                                                      -\kb & 0 & 0 & 0 \\
                                                      0 & 0 & 0 & \ka\kb \\
                                                      0 & 0 &-\ka\kb & 0
                                                   \end{BMAT}
                                                 \right)
                                                 X_2 \\
\phantom{\Phi\left( X_1, X_2 \right)}{} = X_1^T
                                                  \left(
                                                   \begin{BMAT}(e){cc:cc}{cc:cc}
                                                      0 & -1 & 0 & 0 \\
                                                      1 & 0 & 0 & 0 \\
                                                      0 & 0 & 0 & -\ka \\
                                                      0 & 0 & \ka & 0
                                                   \end{BMAT}
                                                 \right)
                                                 X_2.
\end{gather*}
If we let this last matrix describe a (possible degenerate) symplectic form $\varpi$, then $\langle \langle X_1, JX_2 \rangle \rangle = \kb  \varpi \left(  X_1, X_2 \right)$, so that $\varpi$ is compatible with $J$.  In the degenerate case when $\kb = 0$ and the above calculations for $\Phi \left( X_1, X_2 \right)$ make no sense, we can write $\kb \Phi \left( X_1, X_2 \right) = \langle \langle JX_1, X_2 \rangle \rangle$ and we then rescale\footnote{Similarly we rescaled by dividing by $\ka$ or $\kb$, even if they were equal to zero, in order to obtain the metrics on the homogeneous spaces $S^3_{\ka, \kb} / \langle \I \rangle$, $S^3_{\ka, \kb} / \langle \J \rangle$, and $S^3_{\ka, \kb} / \langle \K \rangle$.} by canceling out the factor of $\kb$ on both the left and the right sides of this equation.

The unit sphere $S_{\ka,\kb}^3$ in $\R^4$ can be described by the equation $\langle \langle X, X \rangle \rangle = 1$.  The orbit of $\left( z_1, z_2 \right)$ in $S^3_{\ka, \kb}$ under the $S^1_{\kb}$ action that is given by
\begin{gather*}
\left( z_1, z_2 \right)   \mapsto e^{i \theta} \left( z_1, z_2 \right)
                                   = \left( \CB (\theta) + i \SB (\theta) \right) \left( x + i y, u + i v \right) \\
\phantom{\left( z_1, z_2 \right)}{}  = \big[ \CB (\theta) x - \kb \SB (\theta) y + i \left( \CB (\theta) y + \SB (\theta) x \right),\\
\phantom{\phantom{\left( z_1, z_2 \right)= \quad }{}}{}
                             \CB (\theta) u - \kb \SB (\theta) v + i \left( \CB (\theta) v + \SB (\theta) u \right) \big]
\end{gather*}
is, in real terms, given by $X \mapsto X \CB (\theta) + JX \SB (\theta)$.  Since $X$ and $JX$ are orthogonal\footnote{Neither $X$ nor $JX$ is the zero vector, and we also have that $\langle \langle X, JX \rangle \rangle = \mathcal{R} \left( \langle X, JX \rangle \right) = \mathcal{R}(-i) = 0$.}, a~unit circle is traced out in the plane of $\R^4$ that contains both $X$ and $JX$: Recall that $C^2(\theta) + \kb S^2(\theta) = 1$, noting that $\langle \langle JX, JX \rangle \rangle = \kb$ as can be calculated directly taking into account the fact that $\langle \langle X, X \rangle \rangle = 1$.  This calculation also shows that all f\/ibers are of the same size.

The vector tangent to the f\/iber through $\left( z_1, z_2 \right)$ is given by $i \left( z_1, z_2 \right)$ or, in real terms, by $JX$.  If $X(t)$ is a dif\/ferentiable curve lying in $S^3_{\ka, \kb}$, then $\left| X(t) \right|^2 = 1$ implies that $\langle \langle X(t), \dot{X}(t) \rangle \rangle = 0$.  If this curve is also orthogonal to the f\/iber passing through $X(t)$, then we must also have that $\langle \langle JX(t), \dot{X}(t) \rangle \rangle = 0$.  So $\lambda = \overline{z} dz = \overline{z_1} dz_1 + \ka \overline{z_2}dz_2$ is the principal connection form as $\lambda$ is clearly equivariant.

The curvature form is $d\lambda = d\overline{z_1} \wedge dz_1 + \ka d\overline{z_2} \wedge dz_2$.  Now if $X$ and $Y$ are curves on $S^3_{\ka, \kb}$ so that $\dot{X} = \J q$ and $\dot{Y} = \K q$ at a point $q \in S^3_{\ka, \kb}$, then $d\lambda \left( \dot{X}, \dot{Y} \right) = \langle \dot{X}, \dot{Y} \rangle = \J q \overline{\K q} = -  \J q \overline{q} \K = -\ka \I$.  Also, recall that $\left[ \J, \K \right]  = 2\ka \I$.  And so the curvature is given by $\ka$ and $\lambda$ is a contact form exactly when $\ka \neq 0$.  The area of the inf\/initesimal rectangle $D$ def\/ined by $\epsilon \J q$ and $\delta \K q$ has area $\epsilon \delta$ and the holonomy $\theta$ around the rectangle is given by $\epsilon \delta \ka$:  So the curvature is def\/ined by $\theta = \int_D K \, dA$ or $\epsilon \delta \ka = \epsilon \delta K$ so that $K = \ka$.  Our def\/inition of area is in lieu of rescaling metrics:  See~\cite{HS02} or~\cite{M06} for dif\/ferent calculations of the the curvature.  In ef\/fect we are treating $\{ \I q, \J q, \K q \}$ as an orthonormal frame.

Finally we will use the principal connection form to derive the metric of the spacetime $\Sigma_{\ka, \kb}$ (see also~\cite{M07}):  Recall that the metric is to be rescaled by dividing by $\ka$.  We def\/ine a horizontal curve $X(t)$ (so $\overline{X}(t) \dot{X}(t) = 0$) passing through $q = z_1 + z_2 \J$ in $S^3_{\ka, \kb}$ as follows
\begin{gather*}
X(t)   = \left( \CA (t) + \SA (t) \J \right) \left( z_1 + z_2 \J \right)   = \left( \CA (t) z_1 - \ka \SA (t) \overline{z_2} \right) + \left( \CA (t) z_2 + \SA (t) \overline{z_1} \right) \J.
\end{gather*}
Then
\begin{gather*}
\dot{X}(t)   = \left( -\ka \SA (t) z_1 - \ka \CA (t) \overline{z_2} \right) + \left( -\ka \SA (t) z_2 + \CA (t) \overline{z_1} \right) \J, \\
\dot{X}(0)   = -\ka \overline{z_2} + \overline{z_1} \J, \\
\left| \overline{X}(0) \right|^2  = \ka^2 \left| z_2 \right|^2 + \ka \left| z_1 \right|^2   = 1.
\end{gather*}
Now $X(t)$ is the horizontal lift of the curve $w(t)$ in $\Sigma_{\ka, \kb}$ where
\begin{gather*}
w(t)   = \frac{\CA (t) z_2 + \SA (t) \overline{z_1}}{\CA (t) z_1 - \ka \SA (t) \overline{z_2}}, \\
w(0)   = \frac{z_2}{z_1} = w, \\
\dot{w}(t)   = \left( -\ka \SA (t) z_2 + \CA (t) \overline{z_1} \right)
                                \left( \CA (t) z_1 - \ka \SA (t) \overline{z_2} \right) -
                                \left( \CA (t) z_2 + \SA (t) \overline{z_1} \right)\\
\phantom{\dot{w}(t)   =}{}\times
                                \left( -\ka \SA (t) z_1 - \ka \CA (t) \overline{z_2} \right)
                                \left(  \CA (t) z_1 - \ka \SA (t) \overline{z_2}  \right)^{-2}, \\
\dot{w}(0)   = \frac{\overline{z_1}z_1 + \ka z_2\overline{z_2}}{z_1^2}  = \frac{1}{z_1^2}.
\end{gather*}
Since
\begin{gather*}
\left| z_1 \right|^2 \left( 1 + \ka \left| w \right|^2 \right)   = \left( 1 + \frac{\ka \left| z_2 \right|^2}{\left| z_1 \right|^2} \right) \left| z_1 \right|^2  = \left| z_1 \right|^2 + \ka \left| z_2 \right|^2  = 1,
\end{gather*}
then $\left| z_1 \right|^2 = \frac{1}{\left( 1 + \ka \left| w \right|^2 \right) }$.  As $\left| \J q \right|^2 = \ka$ and $\left| \K q \right|^2 = \kc$, the metric $\frac{1}{\ka}ds^2$ induced on the base space is given by
\[ z_1^2 \overline{z_1}^2 dw d\overline{w} = \frac{dw \, d\overline{w}}{\left( 1 + \ka |w|^2 \right)^2}. \]

\begin{theorem}
Let $H \rightsquigarrow 2\J$, $P \rightsquigarrow 2\K$, and $K \rightsquigarrow 2\I$ denote the respective generators for time translations, space translations, and boosts of the kinematical algebra with commutators
\[ \left[ K, H \right] = P,  \qquad \left[ K, P \right] = -\kb H, \qquad \mbox{and}  \qquad \left[ H, P \right] = \ka K. \]
Then the principal fiber bundle
\[
\begin{CD}
S^3_{\ka, \kb} \\
\pi@VVV  \\
S^3_{\ka, \kb}/\langle \I \rangle
\end{CD}
\]
has $\lambda = \overline{z_1} dz_1 + \ka z_2 \overline{dz_2}$ as its principal connection form.  The base space, which is the space of events, has induced metric
\[ ds^2 = \frac{dw \, d\overline{w}}{\left( 1 + \ka |w|^2 \right)^2} \]
and constant curvature $\ka$.
\end{theorem}

In conclusion, it is hoped that the aims of this paper were met, that the nice structure of the Hopf f\/ibration $S^3 \longrightarrow S^2$ was  generalized in an appealing way, not only for the classical Riemannian surfaces of constant curvature, but especially for the study of $(1 + 1)$ kinematics.  It is also hoped that the reader will f\/ind that these f\/ibrations give a new perspective on  these simple kinematical structures.  Finally, I wish to thank the reviewers for their many helpful suggestions on how this paper could be improved.

\pdfbookmark[1]{References}{ref}
\LastPageEnding

\end{document}